\documentclass[prl,aps,reprint,showpacs,superscriptaddress,amsmath]{revtex4-1}

\usepackage{bm}
\usepackage{graphicx}
\usepackage{dsfont}

\newcommand{\bra}[1]{\langle #1 | \,}
\newcommand{\ket}[1]{\, | #1 \rangle}
\newcommand{\braket}[2]{\langle #1 | #2 \rangle}
\newcommand{\expv}[1]{\langle #1 \rangle}

\newcommand{\Om}{\Omega}

\newcommand{\de}{\delta}
\newcommand{\De}{\Delta}

\newcommand{\Na}{N^{(\mathrm{a})}}
\newcommand{\Np}{N^{(\mathrm{p})}}
\newcommand{\hlf}{\mbox{$\frac{1}{2}$}}
\newcommand{\mc}[1]{\mathcal{#1}}

\begin{document}

\title{Assessing the number of atoms in a Rydberg-blockaded mesoscopic ensemble}

\author{David Petrosyan}
\email{david.petrosyan@iesl.forth.gr}
\affiliation{Institute of Electronic Structure and Laser, 
FORTH, GR-71110 Heraklion, Crete, Greece}
\affiliation{Department of Physics and Astronomy, Aarhus University,
DK-8000 Aarhus C, Denmark}

\author{Georgios M. Nikolopoulos}
\affiliation{Institute of Electronic Structure and Laser, 
FORTH, GR-71110 Heraklion, Crete, Greece}

\date{\today}

\begin{abstract}
The dipole blockade of multiple Rydberg excitations in mesoscopic atomic 
ensembles allows the implementation of various quantum information tasks 
using collective states of cold, trapped atoms. 
Precise coherent manipulations of the collective ground and 
single Rydberg excitation states of an atomic ensemble 
requires the knowledge of the number of atoms with small uncertainty. 
We present an efficient method to acquire such information 
by interrogating the atomic ensemble with resonant pulses 
while monitoring the resulting Rydberg excitations. 
We show that after several such steps accompanied by feedback
the number of atoms in an ensemble can be assessed with high accuracy. 
This will facilitate the realization of high fidelity quantum gates, 
long term storage of quantum information and deterministic sources 
of single photons with Rydberg-blockaded atomic ensembles.
\end{abstract}


\maketitle

%\section{Introduction}

Atoms in high-lying Rydberg states strongly interact with each other 
via the dipole-dipole or van der Waals potential \cite{RydAtoms,rydQIrev}.
This long-range interaction is the basis for many quantum information 
processing schemes \cite{rydQIrev} including quantum gates between 
individual atoms \cite{Jaksch2000,NatPRLSaffman,NatPRLGrangier} and 
photons \cite{GKMFDP0511,Gorshkov201113,Parigi2012,Adams2013}.
The interaction-induced level shifts can suppress resonant optical
excitation of more than one Rydberg atom within a certain blockade
volume containing tens or hundreds of atoms \cite{rydQIrev,rydDBrev}. 
In a seminal paper Lukin {\it et al.} \cite{Lukin2001} proposed to 
employ the transition between the collective ground and single Rydberg 
excitation states of an atomic ensemble to perform various quantum information 
tasks: Collection of atoms confined in a blockade volume can behave as a 
two-level quantum system---a qubit---and can be used for controlled 
manipulation of collective spin excitations as well as on-demand generation 
of single or few photon pulses \cite{Saffman2002,Petrosyan2005}. 
Several related schemes for scalable quantum information processing 
with Rydberg-blockaded atomic ensembles were proposed \cite{rydQIrev} 
while recent experiments have demonstrated (partially) coherent oscillations 
of a single Rydberg excitation \cite{Dudin2012NatPh} and the generation 
of entanglement between a single photon and an atomic ensemble \cite{Li2013}. 

When $N$ atoms are symmetrically (uniformly) excited by a coherent laser field,
the transition amplitude between the collective ground and single-excitation 
states of the ensemble is enhanced by a factor of $\sqrt{N}$ relative 
to the transition amplitude $\Om$ (Rabi frequency) for a single atom. 
Hence, an uncertainty $\De N$ in the atom number translates into 
an uncertainty in the collective Rabi frequency which would result 
in uncontrolled errors during the ensemble manipulations, as detailed below. 
Furthermore, collective Rydberg excitations are typically stored by 
mapping onto long-lived spin excitations of trapped atoms. 
During the storage, the short-range interactions between the atoms 
in different internal states lead to accumulation of a relative 
phase between the collective states which is proportional to the number 
of atoms $N$ and the storage time $\tau$. Hence, again, the atom number 
uncertainty $\De N$ leads to a random relative phase which would reduce 
the storage fidelity or limit its time.

The above arguments attest to the necessity of determining the actual 
number of atoms $\Na$ with sufficient precision in order to achieve 
high-fidelity quantum gates, storage of quantum information and single 
photon generation with a Rydberg-blockaded atomic ensemble. 
Here we present an efficient method to assess $\Na$ with small
uncertainty. The method is based on interrogating the atomic ensemble 
by a sequence of resonant pulses of proper area, each pulse followed 
by a (projective) measurement of Rydberg excitation and a feedback 
to adjust the area of the next pulse. We show that after only 
a few such steps, we gain enough information on the actual number 
of atoms so as to perform quantum gate operations with improved 
fidelity, while after several tens of pulses---and only a few detection 
events---the resulting gate infidelity drops below the threshold for 
fault-tolerant quantum computation \cite{QComp}. 
Moreover, our algorithm performs well even for imperfect detection 
of Rydberg atoms.

%\section{Gate and storage infidelities versus atom number uncertainty}

Before describing the method, let us reexamine resonant laser excitation 
of a collection of $N$ atoms to the strongly interacting Rydberg state. 
We denote by $\ket{g}$ and $\ket{r}$ the ground and Rydberg states
of individual atoms and assume a spatially uniform laser field.
The corresponding atom-field interaction Hamiltonian reads 
$\mc{V}_{\mathrm{af}} =  \hbar \sum_j^N \hlf \Om 
(\ket{r_j}\bra{g_j} + \ket{g_j}\bra{r_j})$. 
The laser field couples the collective ground state 
$\ket{G} = \ket{g_1,g_2,\ldots,g_N}$ to the symmetric single Rydberg excitation
state $\ket{R} = \frac{1}{\sqrt{N}} \sum_j^N \ket{g_1,\ldots,r_j,\ldots, g_N}$
with the collective Rabi frequency 
$\frac{2}{\hbar} \bra{R} \mc{V}_{\mathrm{af}} \ket{G} = \sqrt{N}\Om$. 
Applying a laser pulse with the single-atom pulse area 
$\theta = \int \Om \, dt$ transforms the initial state $\ket{G}$ as
\begin{equation} 
\ket{G} \to \cos \big(\hlf \sqrt{N} \theta \big) \ket{G} 
+ i \sin \big( \hlf \sqrt{N} \theta \big) \ket{R} , \label{eq:Gtrnsf}
\end{equation}
and similarly for the initial $\ket{R}$ (with the replacement 
$\ket{G} \leftrightarrow \ket{R}$).
Throughout this paper, we assume that multiple Rydberg excitations
are strongly suppressed by the dipole blockade 
\cite{rydQIrev,rydDBrev,Lukin2001,Dudin2012NatPh}.

We can estimate the average fidelity of performing quantum gates 
with Rydberg-blockaded atomic ensembles in a realistic experiment. 
Usually, upon preparing an ensemble of atoms in a trap, the experimentalist 
does not know their actual number, but he can assign to it a random variable 
$N$ with a certain probability distribution $P(N)$ and a mean $\expv{N}$.
It is then natural to take as the most probable number of atoms 
$\Np = \expv{N}$. Consider, e.g., the \textsc{not} (or \textsc{swap}) 
gate which swaps the states $\ket{G}$ and $\ket{R}$.
To within a trivial ($\frac{\pi}{2}$) phase shift of the final state, 
the \textsc{swap} is realized by applying a collective $\pi$-pulse 
$\sqrt{\Np} \theta = \pi$. If the actual number of atoms $\Na = \expv{N}$, then 
$\ket{G} \to i \ket{R}$ and $\ket{R} \to i \ket{G}$ [cf. Eq.~(\ref{eq:Gtrnsf})].
But in general the number of atoms is uncertain with the standard 
deviation $\De N$. Defining the fidelity as 
$F \equiv |\braket{\Psi^{(\mathrm{t})}}{\Psi^{(\mathrm{a})}}|$,
where $\ket{\Psi^{(\mathrm{t})}}$ is the ideal target state and 
$\ket{\Psi^{(\mathrm{a})}}$ is the actually obtained one, we have 
\begin{equation} 
F = \left| \sin \left(\frac{\pi}{2} \sqrt{\frac{\Na}{\Np} } \right) \right| . 
\label{eq:Fid}
\end{equation}  
To estimate the average fidelity, we may replace $\Na \to \expv{N} \pm \De N$
assuming $\De N \ll \expv{N}$, which leads to  
$\bar{F} \approx \cos \left( \frac{\pi \De N}{4 \expv{N}} \right)$.
Then the average infidelity, i.e., deviation of $\bar{F}$ from unity, scales 
as $1-\bar{F} \approx \frac{\pi^2}{32} \left(\frac{\De N}{\expv{N}}\right)^2$.
It is easy to verify that the infidelity for any single-qubit rotation
gate scales quadratically with the uncertainty $\De N$.
Typically, the probability distribution of the atom number is Poissonian 
$P(N) = P_{\mathrm{Poisson}}(N) \equiv \expv{N}^N e^{-\expv{N}}/N!$ 
with the variance $(\De N)^2 = \expv{N^2} - \expv{N}^2 = \expv{N}$.
Then the average infidelity reduces to 
$1-\bar{F} \approx \frac{\pi^2}{32 \expv{N}}$
and to have a small error of, e.g., $1-\bar{F} \lesssim 10^{-4}$ requires  
$\expv{N} \gtrsim 3000$ atoms, which is an impractically large number
for few $\mu$m sized trap (to guarantee the blockade). 

Next, once a Rydberg excitation $\ket{r}$ is created, it is transferred 
to a lower spin-flip state $\ket{s}$, which maps the collective 
state $\ket{R}$ onto the metastable state 
$\ket{S} = \frac{1}{\sqrt{N}} \sum_j \ket{g,\ldots,s_j,\ldots, g}$. 
Cold atoms in a tight trap interact with each other, albeit weakly,
with the strength $U_{\mu \nu}$ ($\mu ,\nu = g,s$) determined by the
(state-dependent) $s$-wave scattering length and trap geometry \cite{UCgases}.
The interaction energies of states $\ket{G}$ and $\ket{S}$ are then given 
by $E_G = \hlf U_{gg} N(N-1)$ and  $E_S = \hlf U_{gg} (N-1)(N-2) + U_{gs}(N-1)$, 
and their difference is $E_G - E_S = \de U(N-1)$, where 
$\de U \equiv U_{gg} - U_{gs}$.

We now estimate the average fidelity of storage of a quantum state 
in a coherent superposition of $\ket{G}$ and $\ket{S}$. 
Unless $U_{gg} = U_{gs}$ ($\de U =0$), the uncertainly $\De N$ 
in the total atom number results in the accumulation during time $\tau$ 
of a random relative phase $\De \phi = \De N \de U \tau$ between 
states $\ket{G}$ and $\ket{S}$. Upon averaging over all qubit states,
the storage fidelity is $\bar{F} = \sqrt{[1+\cos(\De \phi)]/2}$, and for 
$\De \phi \ll 1$ the infidelity $1-\bar{F} \approx (\De \phi)^2/8$ 
scales again quadratically with the uncertainty $\De N$. 
For the Poissonian distribution of the atom number, we then have 
$1-\bar{F} \approx \expv{N} (\de U \tau)^2/8$, i.e., small error for 
longer storage times requires now a smaller mean number of atoms $\expv{N}$.
As an example, with $\expv{N} \simeq 200$ and $\de U /2\pi =1-10\:$Hz,
the infidelity $1-\bar{F} \lesssim 10^{-4}$ limits the storage time to
$\tau \lesssim 0.1 \:$ms.

%\section{Description of the algorithm}

Having discussed the detrimental effects of the atom number uncertainty, 
we now turn to the description of a protocol to determine  with high accuracy
the actual number of atoms $\Na$. This is done in a finite number of steps, 
each step consisting of 
(i) deducing the most probable number of atoms $\Np$ 
from the probability distribution $P(N)$, 
(ii) applying to the atomic ensemble in state $\ket{G}$ a collective pulse 
of area $\sqrt{\Np} \theta = 2\pi m$ ($m \in \mathds{Z}$), 
(iii) measuring the Rydberg excitation 
and updating $P(N)$ using the Bayes' rule. 
Specifically, before each step $i$, there is a probability 
distribution $P_{i-1}(N)$ of the atom number which reflects our current 
state of knowledge---or ignorance, for that matter---about the actual 
number of atoms. As the most probable atom number, we take $\Np_{i-1}$ 
which maximizes $P_{i-1}(N)$. We then set the collective pulse area 
$\sqrt{\Np_{i-1}} \theta_i$ to be a multiple ($m$) of $2\pi$, 
i.e., $\theta_i = 2 m \pi/\sqrt{\Np_{i-1}}$. 
After the pulse, we perform projective measurement of the Rydberg 
excitation of the ensemble, which may be accompanied by either 
discarding the atom in state $\ket{r}$ or recycling it back to 
state $\ket{g}$. If the result of the measurement is negative, 
the updated probability distribution becomes 
$P_{i}(N) = \tilde{\sigma}_i(N) \, P_{i-1}(N) / S_i \; \forall \; N$,
where $\tilde{\sigma}_i(N) = \cos^2(\hlf \sqrt{N} \theta_i)$ is the 
probability of no Rydberg excitation [cf. Eq.~(\ref{eq:Gtrnsf})]; 
if the measurement does yield a Rydberg excitation, 
then the updated probability distribution is 
$P_{i}(N) = \sigma_i(N) \, P_{i-1}(N) / S_i \; \forall \; N$ 
(if the Rydberg atom is transferred back to state $\ket{g}$ 
\cite{Dudin2012NatPh}) or 
$P_{i}(N) = \sigma_i(N+1) \, P_{i-1}(N+1) / S_i  \; \forall \; N$
(if the Rydberg atom is removed \cite{Schwarzkopf2011}), 
where $\sigma_i(N) = \sin^2 \big( \hlf \sqrt{N} \theta_i \big)$ is the 
probability of Rydberg excitation [Eq.~(\ref{eq:Gtrnsf})] 
and $S_i$ renormalizes $P_{i}(N)$.  
In turn, in the (numerical or laboratory) experiment, the outcome 
of the measurement is determined by the actual---but unknown---number 
of atoms $\Na$ through the actual probability of Rydberg excitation
\begin{equation}
\sigma_{i}^{(\mathrm{a})} = \sin^2 \left(m\pi \sqrt{\frac{\Na}{\Np_{i-1}}} \right)
 . \label{eq:siga}
\end{equation}

To commence the procedure, we assume a certain mean number of atoms 
$\expv{N}$ and a width $W$ of the atom number distribution $P_0(N)$. 
For definiteness, we take as a seed the Poissonian distribution 
$P_0(N) = P_{\mathrm{Poisson}}(N)$ (with FWHM $W = 2 \sqrt{2 \ln(2)\expv{N}}$),
but any other distribution with a well-defined range (and no long wings) 
will do as well. For the first step, the most probable atom number 
is $\Np_0 = \expv{N}$. The positive integer constant $m$ is then 
chosen so as to maximize the detection probability $\sigma(N)$ 
at the edges of the initial distribution, e.g., at 
$N \simeq \expv{N} \pm \frac{3}{2} W$, which leads to 
$m = \lfloor 2 \expv{N}/3W \rfloor = \lfloor \sqrt{\expv{N}}/3.53 \rfloor$,
where $\lfloor \ldots \rfloor$ denotes the floor function.  

In the adaptive (feedback) scheme described above, no Rydberg excitation 
at a given step $i$ narrows the probability distribution $P_{i}(N)$ around 
the perceived atom number $\Np_{i-1}$, while a detection of Rydberg 
excitation drastically reshapes $P_{i}(N)$ by opening a hole around 
$\Np_{i-1}$ which leads to a new $\Np_i$ for the following step. 
Observe now that the actual probability of Rydberg excitation in 
Eq.~(\ref{eq:siga}) is appreciable when the perceived $\Np$ and actual 
$\Na$ atom numbers differ significantly, while $\sigma^{(\mathrm{a})}$ is small 
when $\Np \simeq \Na$, and $\sigma^{(\mathrm{a})} = 0$ and remains so 
if our guess is correct, $\Np = \Na$. Hence, if the initial difference 
$\Np_0 - \Na$ is large, $\Np$ will rapidly change during the first several 
steps, but once it is close to $\Na$, its further approach to $\Na$ will
slow down. One may optimize the search of the atom number by readjusting 
$m$ at each step, but for simplicity we fix $m$ from the beginning,
which still leads to a rapid improvement of fidelity, as attested below.  

%%%%%%%%%%%%%%%%FIGURE%%%%%%%%%%%%%%%%
\begin{figure}[t]
\includegraphics[width=8.5cm]{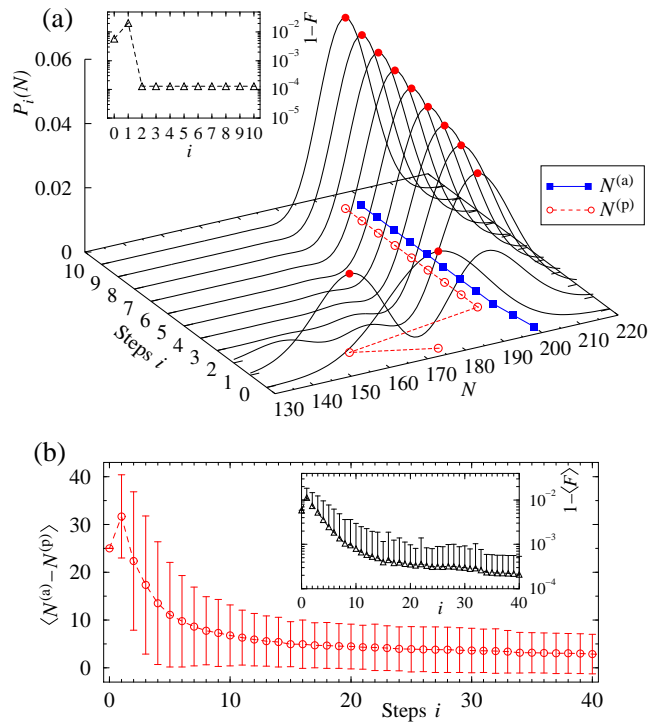}
\caption{%(color online).
(a) Single realization of the algorithm for the initial 
mean $\expv{N} = 175$ and actual $\Na = 200$ number of atoms:  
Main graph shows the probability distribution $P_i(N)$ and 
the deduced atom number $\Np_i$ after $i$ steps
($\Na$ decreases by 1 after each Rydberg detection event---steps $i=1,2$), 
and the inset shows the corresponding infidelity $1-F$ 
of the \textsc{swap} gate [cf. Eq. (\ref{eq:Fid})].
(b) Many ($10^3$) independent realizations of the algorithm 
for the same initial conditions:
Main graph shows the mean value $\expv{\Na - \Np}$ of the difference 
between the actual and deduced atom numbers after $i$ steps, and 
the inset shows the corresponding mean infidelity $1-\expv{F}$ 
(error bars are one standard deviation).}
\label{fig:strj}
\end{figure}
%%%%%%%%%%%%%%%%%%%%%%%%%%%%%%%%%%%%%%%

To demonstrate the algorithm, in Fig.~\ref{fig:strj}(a) we show the results
of a single trajectory simulation for several steps of the stochastic 
process. In the numerical experiment, the occurrence of Rydberg 
excitations and their detection is determined via Monte Carlo
procedure. Namely, at every step we draw a uniform random number 
$q_{\sigma} \in [0,1]$ and compare it with the current probability 
$\sigma^{(\mathrm{a})}$: if $q_{\sigma} < \sigma^{(\mathrm{a})}$ the
ensemble is in the excited state $\ket{R}$, otherwise it is in the
ground state $\ket{G}$. Assuming for now a perfect detector, whenever 
a Rydberg atom is produced, it is detected and removed from the ensemble, 
which ends up in state $\ket{G}$ but with one atom less, $\Na \to \Na-1$. 
The algorithm works equally well if the Rydberg atom $\ket{r}$ is recycled 
back to the ground state $\ket{g}$, which leaves $\Na$ unchanged. 
But since we deal with $\Na \gg 1$ and create only a few excitations, 
the difference between the two approaches is inconsequential.
In practice, detecting the Rydberg atom via state-selective ionization 
accompanied by its removal is perhaps experimentally easier 
\cite{Schwarzkopf2011} and it does not heat the atomic ensemble.

Our benchmark is the infidelity $1-F$ of the \textsc{swap} gate on 
the transition $\ket{G} \to \ket{R}$ executed with the perceived atom 
number $\Np$ [cf. Eq.~(\ref{eq:Fid})]. This is shown in the inset of 
Fig.~\ref{fig:strj}(a). Clearly, once $\Np$ is close to the actual 
atom number $\Na$, the infidelity is very small. 

In Fig.~\ref{fig:strj}(b) we show the mean difference $\expv{\Na - \Np}$
between the actual $\Na$ and deduced $\Np$ number of atoms after $i$ steps,
as obtained from many independent realizations of the algorithm. 
With increasing $i$, this difference and its statistical dispersion 
are decreasing, while on average less than three Rydberg excitation 
events occur by $i=40$, mainly during the first several steps. 
The corresponding mean infidelity $1-\expv{F}$ of the \textsc{swap} 
gate drops from the initial $1-\expv{F} \gtrsim 10^{-2}$ 
to $2 \times 10^{-4}$ with the uncertainly $\De F \simeq 3 \times 10^{-4}$.

%\section{Analysis of the perfomance}

%%%%%%%%%%%%%%%%FIGURE%%%%%%%%%%%%%%%%
\begin{figure}[t]
\includegraphics[width=8.7cm]{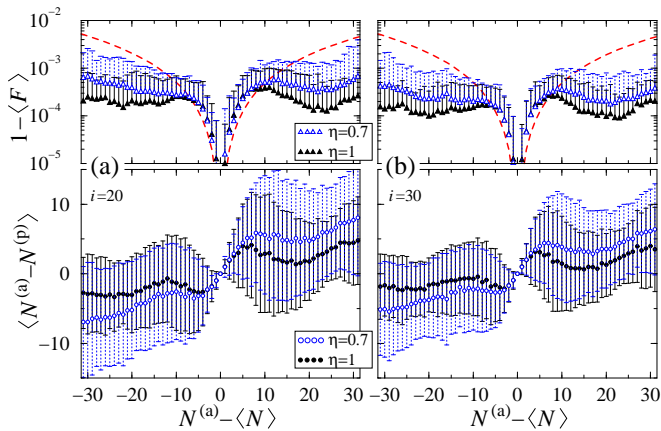}
\caption{%(color online).
Many independent realizations of the algorithm for the initial mean number 
of atoms $\expv{N} = 250$ after 
(a) $i=20$ and (b) $i=30$ steps, with the Rydberg excitation detection
efficiency $\eta = 1$ (black, solid symbols) and $\eta = 0.7$ 
(blue, open symbols). The mean values $\expv{\Na - \Np}$ of the 
difference between the actual $\Na$ and deduced $\Np$ atom numbers (bottom) 
and the corresponding infidelity $1-\expv{F}$ of the \textsc{swap} gate (top)
are plotted versus the initial $\Na$ (error bars are one standard deviation).
Red dashed line is the initial infidelity 
[cf. Eq. (\ref{eq:Fid}) with $\Np = \expv{N}$].}
\label{fig:NNF}
\end{figure}
%%%%%%%%%%%%%%%%%%%%%%%%%%%%%%%%%%%%%%%

In a laboratory experiment, the detector measuring the Rydberg excitations 
is never perfect---the best detection efficiency of individual Rydberg 
atoms achieved so far is $\eta \simeq 0.75$ \cite{Schauss2012}.
In our numerical simulations, we can account for finite detection 
efficiency $\eta \leq 1$ by reducing the probability of detecting 
the Rydberg excitation by the corresponding factor; 
at each step, however, the probability of producing a Rydberg excitation 
is still given by $\sigma_{i}^{(\mathrm{a})}$. More precisely, once 
a Rydberg excitation is produced ($q_{\sigma} < \sigma^{(\mathrm{a})}$), 
we draw another random number $q_{\eta} \in [0,1]$ and compare it to $\eta$:
if $q_{\eta} < \eta $ the Rydberg atom is detected; otherwise it is not 
and we proceed as if it were not created, but still assume that 
any Rydberg atom is removed from the ensemble after each step.

\begin{figure}[t]
\includegraphics[width=5.5cm]{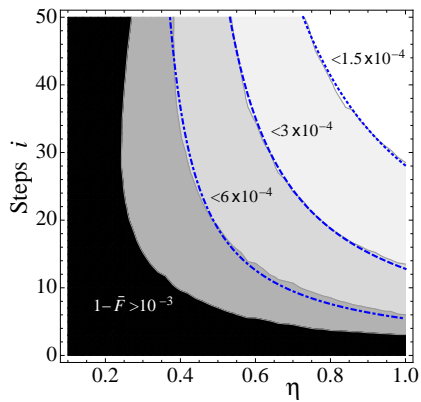}
\caption{%(color online).
Average infidelity $1-\bar{F}$ of the \textsc{swap} gate 
versus the detector efficiency $\eta$ and the number of steps. 
The isolines at $(1.5;3;6) \times 10^{-4}$ are approximated 
by function $f(\eta) = \frac{a}{\eta + b}$ with 
$a,b = (17.45,-0.378;8.056,-0.341;3.84;-0.295)$, respectively
(blue dotted; dashed; dot-dashed lines).}
\label{fig:inFetai}
\end{figure}

In Fig.~\ref{fig:NNF} we show the mean difference $\expv{\Na - \Np}$
and the corresponding infidelity $1-\expv{F}$ of the \textsc{swap} gate,
versus the initial actual atom number $\Na$, resulting after
$i=20$ and $i=30$ steps. Apparently, $\eta < 1$ does not invalidate 
our approach, since increasing the number of steps can compensate for 
the loss of information effected by imperfect measurements. To verify
this point, in Fig.~\ref{fig:inFetai} we show the average infidelity
$1- \bar{F} \equiv \sum_{\Na} P_{\mathrm{Poisson}}(\Na) \, (1 - \expv{F})$
as a function of the detector efficiency $\eta$ and the number of steps $i$.
Attaining larger fidelity using a detector with smaller efficiency
requires increasing the number of steps $i \propto 1/\eta$.

%%%%%%%%%%%%%%%%FIGURE%%%%%%%%%%%%%%%%
%%%%%%%%%%%%%%%%%%%%%%%%%%%%%%%%%%%%%%%

%\section{Concluding remarks}

Before closing, let us briefly survey the practical side of our work. 
In the current experiments \cite{Dudin2012NatPh,Li2013}, the uncertainty
in the atom number $\De N$ is perhaps a lesser evil than an imperfect 
blockade of multiple Rydberg excitations. The latter problem can be 
overcome by using higher-lying (stronger interacting) Rydberg states 
and tighter confinement of the atoms. Single Rydberg excitations 
transferred to a lower metastable state can then be used for on-demand
generation of single photons, which is accomplished by stimulated Raman 
emission requiring sizable optical depth (OD) of the ensemble \cite{stRamOpt}.
In a blockade volume of several $\mu$m size, $\expv{N} \gtrsim 200$ 
atoms are needed for $\mathrm{OD} \gtrsim 5$. Larger $\expv{N}$ is better, 
both for attaining larger OD and smaller relative uncertainty 
$\frac{\De N}{\expv{N}} = \expv{N}^{-1/2}$, but too high atom densities 
$\rho_{\mathrm{at}} \gtrsim 10^{13-14} \:$cm$^{-3}$ usually lead to 
excessive collisional decoherence and losses due to the three-body
recombination. Hence, we are constrained to deal with the ensembles of 
a few hundred atoms $\expv{N}$ and large associated uncertainty $\De N$. 

Our method to accurately assess the number of atoms is quite efficient, 
requiring $i=20-30$ resonant pulses and measurements of Rydberg excitations. 
With the typical single-atom Rabi frequency $\Om$ in the MHz range, 
each pulse duration is $t_\mathrm{p} \lesssim 1\:\mu$s, while the Rydberg
atom detection through ionization would take $t_\mathrm{d} \sim 10\:\mu$s.
The complete protocol can then be accomplished in less than a ms, 
yielding the actual number of atoms $\Na$ with a small uncertainty 
$\De N < \sqrt{\Na}$. We note that small variations 
in the single-atom pulse area $\Delta \theta < \frac{\De N}{2\Na}$, 
due to uncertainty in $\Om$ or $t_\mathrm{p}$, do not significantly 
affect the performance of our algorithm and the fidelity of subsequent
quantum gates.

The algorithm we have used to deduce the actual number of atoms is not 
unique---we have explored other, somewhat less efficient methods too. 
Yet, the general concept of interrogating the atomic ensemble by 
appropriate pulses in order to extract the atom number with reduced 
uncertainty applies to other approaches as well.

To summarize, we have exposed the detrimental effects of the atom number 
uncertainty on the fidelity of quantum gates with the Rydberg-blockaded 
atomic ensembles. We have proposed and analyzed an experimentally realistic
method to reduce this uncertainty and thereby increase the fidelity of 
collective gate operations and storage of quantum information in mesoscopic
atomic ensembles. Accurate determination of the atom number will also facilitate
deterministic generation of single photons \cite{Saffman2002,Petrosyan2005}
and extraction of single Rydberg atoms or ions \cite{Ates2013} from 
mesoscopic atomic ensembles. We note the recent proposals
\cite{Beretov201113,DPKM2013} employing adiabatic passage to reliably 
accomplish the above tasks in few-atom systems. Preparing number-squeezed 
clouds of cold atoms \cite{Hume2013} may find further applications 
in precision metrology and studies of Bose-Einstein condensation.

\begin{acknowledgments}
We thank I. Lesanovsky and K. M\o lmer for helpful comments and suggestions.
\end{acknowledgments}

\end{document}